\documentclass[conference,a4paper]{APSIPA2023}
\usepackage{multirow}
\usepackage{graphicx}
\usepackage{amsmath}
\usepackage[psamsfonts]{amssymb}
\usepackage{amsxtra}
\usepackage{threeparttable}
\usepackage{cite}
\usepackage{booktabs}

\usepackage{fancyhdr}

\fancypagestyle{firststyle} {
    \fancyhf{}
    \fancyhead[C]{2023 Asia Pacific Signal and Information Processing Association Annual Summit and Conference (APSIPA ASC)}
}

\begin{document}

\title{Channel-Spatial-Based Few-Shot Bird Sound Event Detection}
\author{
    Lingwen Liu, Yuxuan Feng, Haitao Fu, Yajie Yang, Xin Pan, and Chenlei Jin \\
    Jilin Agricultural University, Changchun, Jilin, China  \\
    LingWeasley@outlook.com, fengyuxuan.cn@163.com
}

\maketitle
\thispagestyle{firststyle}
\pagestyle{fancy}

\begin{abstract}
In this paper, we propose a model for bird sound event detection that focuses on a small number of training samples within the everyday long-tail distribution. As a result, we investigate bird sound detection using the few-shot learning paradigm. By integrating channel and spatial attention mechanisms, improved feature representations can be learned from few-shot training datasets. We develop a Metric Channel-Spatial Network model by incorporating a Channel Spatial Squeeze-Excitation block into the prototype network, combining it with these attention mechanisms. We evaluate the Metric Channel Spatial Network model on the DCASE 2022 Take5 dataset benchmark, achieving an F-measure of $66.84\%$ and a PSDS of $58.98\%$. Our experiment demonstrates that the combination of channel and spatial attention mechanisms effectively enhances the performance of bird sound classification and detection.
\end{abstract}

\section{Introduction}
Birdsong, or the sounds produced by birds, serves as a means of communication that enables birds to interact with one another. Through their calls, birds can recognize each other's species, gender, age, territory, and mate information. Moreover, these sounds can function as warnings or indicate the location of food sources. Each bird species has its own unique calls, distinguishable by their frequency, pitch, duration, and rhythm. Some bird sounds are exceptionally beautiful, even considered a form of music. Studying bird sounds allows researchers to gain a better understanding of bird behavior~\cite{Brumm}. Many scientists and bird enthusiasts utilize bird sounds to identify species, track migration patterns, and examine the ecological environment~\cite{Bateman}.

Birds are prevalent in various ecosystems, and detecting the status of their sounds can offer insights into the diversity and abundance of bird species in a given area. Shifts in the frequency and pitch of bird sounds may signal threats or harm to the ecosystem, such as pollution, habitat loss, or climate change. Consequently, bird calls can serve as indicators of ecosystem health. By monitoring and analyzing bird sounds, we can evaluate the health of an ecosystem and take appropriate measures to protect and restore it~\cite{Stowell}. Birds help maintain the ecological balance in the natural world, enhance the environment's naturalness, and serve as valuable resources for teaching and research. Monitoring the natural environment through bird sound detection can directly benefit humans, underscoring the importance of bird sound detection in ecological monitoring.

Presently, various methods are employed in bird sound event detection tasks~\cite{Mesaros, Yang}, including template matching~\cite{Morfi}, multiple-instance learning~\cite{maron1997framework,carbonneau2018multiple,briggs2012rank}, and prototype networks~\cite{Snell, Zhang, Anderson}. Template matching is an algorithm that utilizes mathematical models such as Gaussian Mixture Models~\cite{yang1998gaussian} or Hidden Markov Models~\cite{khademi2017hidden}. In bird sound event detection tasks, this method employs fast and normalized cross-correlation to identify templates within spectrograms. Subsequently, the template matching technique calculates the cross-correlation between each event in the dataset and the remaining audio data, setting a distinct threshold for each file based on the maximum value of the cross-correlation results.

Template matching was the earliest method used for bird sound event detection, but because it can only use translational matching during matching, the feature matching of the template matching method is not complete enough. In addition, multiple-instance learning is also used, assuming that a piece of audio is a package, and each event in the audio is treated as an instance without labels. The positive or negative of the event is judged based on the labels of these packages. If a package is marked as a positive package, there is at least one positive instance in the package. Conversely, if a package is marked as a negative package, all instances are negative. The purpose of multiple-instance learning is to make accurate judgments about new packages by learning from packages, which can improve the training results of weakly labeled datasets under the condition of using a large amount of data for training. However, the natural conditions of the bird sound dataset affect the dataset size~\cite{Morfi, Gemmeke}, and using multiple-instance learning methods to process bird sounds-related issues cannot achieve high performance.

On the other hand, a prototype network is a simple and efficient neural network for few-shot learning~\cite{Ravi, Wang,wang2020generalizing}. Its core idea is to create a prototype representation for each category and then classify them through distance calculation, which can achieve higher generalization performance on the same dataset. In terms of distance calculation, we chose metric learning~\cite{Kulis}. In metric learning, different data points are mapped to a feature space, and the distance or similarity between these data points is defined by the metric function. Common metric learning methods include Euclidean distance, Cosine similarity, Mahalanobis distance, Manifold, etc.~\cite{Rodr}. Metric learning has wide applications in fields such as image processing, Natural Language Processing, and Recommendation Systems.

To effectively detect bird sounds in small datasets, we propose a novel prototype network-based approach~\cite{Liu1, Tang} and introduce the Channel Spatial SE module~\cite{Lan} to enhance the dependency between channels and spatial dimensions. Our design, the Metric Channel-Spatial Network (MCS-Net) model, emphasizes channel and spatial features during the feature extraction process. We evaluate the MCS-Net model on the DCASE 2022 Take5 benchmark dataset, achieving an F-measure of $66.84\%$ and a PSDS of $58.98\%$, which surpasses previous works. This result demonstrates the MCS-Net model's effectiveness in improving bird sound detection performance.

The remaining parts of this paper are organized as follows. In Section 2, we introduced the system overview and Channel-Spatial SE. In Section 3, we conducted experiments to demonstrate the performances of the Metric Channel-Spatial Network (MCS-Net) model. In Section 4, we summarize this paper.

\section{Method}
\subsection{Input}
This paper performs bird sounds detection in the frequency domain using the audio data from the DCASE 2022 Take5 dataset. Since the length of these audio files varies, we first need to crop them to a fixed length. Then, We first use the Short-time Fourier Transform (STFT) to convert the audio from the time domain signal to the frequency domain signal and then convert it into Mel Frequency Cepstral Coefficients (MFCC)~\cite{Hossan}, which is a commonly used feature in the field of speech and audio processing and is more suitable for simulating the perceptual characteristics of the human ear. To compress the dynamic range of MFCC and perform automatic gain control, we further preprocess MFCC using Per-Channel Energy Normalization (PCEN)~\cite{Wang1}, which reduces the difference in foreground loudness and suppresses background noise. The inputs of our model are processed by MFCC ($F_{\text{MFCC}} ( \cdot )$) and PCEN ($F_{\text{PCEN}} ( \cdot  )$) respectively, and then concatenated together, which is represented as follows:
\begin{figure}[!h]
\centering
\includegraphics[width=\linewidth]{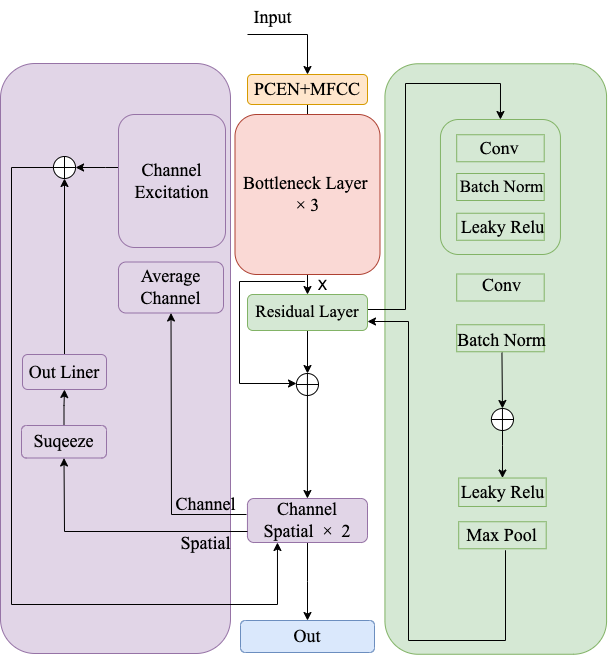}
\caption{ The model structure consists of PCEN+MFCC module, bottleneck layer, BC Residual module, and Channel Spatial module. where orange represents the PCEN+MFCC overlay module, pink represents the prototype network module, green represents the residuals module, and purple represents the Channel Spatial module.}
\label{model_structure}
\end{figure}
\begin{equation}
    F_\text{ALL} (x) = F_\text{MFCC} (x) + F_\text{PCEN} (x)
\end{equation}
where $x$ represents the amplitude spectrogram of the audio data to be detected. We treat $F_{\text{ALL}}$ as the input of the bottleneck layer ($BL$), which is a convolution-based feature extraction block. After processing $F_{\text{ALL}}$ through three $BL$ blocks, we obtain the initial feature map $F_{\text{initial}}$ for each audio segment. To further enhance feature extraction and improve performance, we added an improved Channel Spatial SE block, which adopts the idea of Squeeze-and-Excitation~\cite{Hu} and performs ``squeeze" and ``excitation" from both channel and spatial perspectives, as shown in the purple shaded area in Figure 1. Before squeezing, we first calculate a location value for each channel using the following equation~\ref{eq:2}, where the  value is the statistical number of channel ranges generated by shrinking u from the spatial dimension
\begin{equation}
    G_\text{K} = \frac{1}{H\times W}\Sigma _{i}^{H}\Sigma _{j}^{W}{O_{K}}^{\left ( i,j\right )}
    \label{eq:2}
\end{equation}
\begin{equation}
     \widehat{G} = W_{1}\left ( \sigma \left ( {W_{2}}^{Z}\right )\right ),W_{1}\in R^{C\times \frac{C}{P}},W_{2}\in R^{C\times \frac{C}{P}}
     \label{eq:3}
\end{equation}

In Equation~\ref{eq:3}, $p$ represents the parameter rate, $W_1$ and $W_2$ represent the weights of the two fully connected layers, and $\widehat{G}$ represents the weighted calculation result of the $i_{th}$ channel.
When we input $F_{\text{initial}}$ into Channel Spatial SE, $F_{\text{initial}}$ enters both the channel and spatial branches. When we input $F_{\text{initial}}$ from the channel perspective, first, we obtain a unique feature value $O$ for each channel by global average pooling. Then, we use two fully connected layers with different weights and activate them with the Leaky ReLU activation function. Finally, we normalize the values through the sigmoid layer.
\begin{equation}
\begin{aligned}
       {\widehat{O}_{C}}^{SE} &= {\widehat{F}_{C}}^{SE}\left ( O\right )\\
    &= \left [ \sigma \left ( \widehat{C}_{1}\right )O_{1}, \cdots,\sigma \left ( \widehat{C}_{2}\right )O_{2}, \cdots,\sigma \left ( \widehat{C}_{c}\right )O_{c}\right ]
\end{aligned}
\end{equation}

When we input $F_{\text{initial}}$ from the spatial perspective, we first use a  $1\times 1 \times 1 $ convolutional kernel to reduce the dimension, then perform nonlinear operations with the sigmoid function to obtain the output in the range of [0-1], and finally multiply the output with the spatial dimension of the initial feature map to obtain $\widehat{O}$. The specific calculation equation is as follows:
\begin{equation}
\begin{aligned}
     \widehat{O}&= {{F_{S}}^{SE}}\left ( O\right )\\
     &= \left [ \sigma \left ( q_{1,1}\right )O^{1,1}, \cdots, \sigma \left ( q_{i,j}, \right )O^{i,j}, \cdots,\sigma \left ( q_{H,W}\right )H,W\right ]
\end{aligned}
\end{equation}

Finally, we add the results of the channel and spatial processing together as the final output:
\begin{equation}
    {{\widehat{O}_{SC}}^{SE}} = {{\widehat{O}_{S}}^{SE}} + {{\widehat{O}_{C}}^{SE}}
\end{equation}

We optimize the model using the prototype network and metric learning method. We first calculate a class prototype by embedding several samples into the latent space, and then use the metric learning equation to calculate the distance between the final output and the prototype to determine which class it belongs to. The equation is as follows:
\begin{equation}
d_{2j}^{i}= \sqrt{\sum{(x_{i}^{q}-x_{j}^{s})^2}},
d_{2j+1}^{i}= \sqrt{\sum{(x_{i}^{q}-\widetilde{x}_{j}^{s})^2}}
\end{equation}
\begin{table*}
\centering
\caption{Comparison with Byte Dance in DCASE 2022 Take5. Pre stands for accuracy, Rec stands for recall, F-measure stands for the balance value of accuracy and recall, and PSDS stands for Polyphonic Sound Detection Score, where Templane Matching and ProtoNet do not have PSDS values, and all metrics in the table are reported as percentages.}
\begin{tabular}{ccccc}
\toprule
\textbf{Method}& \textbf{Precision} (\%) & \textbf{Recall} (\%) & \textbf{F-measure(\%)}& \textbf{PSDS(\%)}\\ \midrule
{\color[HTML]{333333} \textbf{Templane Matching}} & 2.42         & 18.32        & 4.28                         & N/A                          \\ \midrule
{\color[HTML]{333333} \textbf{ProtoNet}}          & 36.34        & 24.96        & 29.59                        & N/A                          \\ \midrule
{\color[HTML]{333333} \textbf{DCASE-ByteDance}}   & 69.3         & 57.3         & {\color[HTML]{FE0000} 62.73} & {\color[HTML]{FE0000} 57.52} \\ \midrule
{\color[HTML]{333333} \textbf{MCS-Net}}           & 68.5         & 66.5         & {\color[HTML]{FE0000} 66.84} & {\color[HTML]{FE0000} 58.98} \\
\bottomrule
\end{tabular}
\label{table:1}
\end{table*}
\subsection{Learning Objective}
As this paper is based on a CNN network~\cite{Gong,yao2020comprehensive}, traditional CNNs treat each convolutional layer equally, which ignores the importance of channels and spatial dimensions. Therefore, to increase the attention to channels and spatial dimensions, we added the Channel Spatial SE module. The core idea of the Channel Spatial SE module is to compress the feature map of each channel into a single number (i.e. "squeeze"), and then use a subnetwork called the "excitation" module to learn the weight of each channel. This idea is applied from two different perspectives to dynamically adjust the weight of each channel and improve the representation ability of the feature map, while also increasing the dependency between channels and spatial dimensions, thereby improving the performance of the CNN.

\subsection{Model Structure}

Fig.~\ref{model_structure} shows the structure of the MCS-Net model, which consists of 3 BL modules, 1 PL module, and 2 Channel Spatial SE modules. The BL module consists of a 2D convolutional layer, batch normalization, and a residual block, using the Leaky ReLU activation function in the residual block. The PL module consists of a 1D convolutional layer, batch normalization, and the Leaky ReLU activation function, with an additional average pooling layer at the end. The Channel Spatial SE module consists of two branches, one for channels and one for spatial dimensions. Both branches have the same structure, first using a $1 \times 1 \times 1$ convolutional kernel to "squeeze" the feature representation into a single number, then weighting the channel and spatial features, and finally outputting through convolution and merging the features from both perspectives. Our experiments have demonstrated the importance of features in both channels and spatial dimensions. We added the Channel Spatial SE module to the CNN network to enhance the model's dependency on channels and spatial dimensions, and improved the accuracy of the model through recalibration of the feature map during the learning process.

\section{Experiment}
\subsection{Experimental Data}
The experiments in this paper used the public dataset of task 5 of DCASE 2022. 
The dataset is pre-split into training and validation sets. The training set consists of five sub-folders (BV, HV, JD, MT, and WMW) deriving from a different source each. The number of audio recordings in the  training set is 174, corresponding to 21 hours. Along with the audio files, multi-class annotations are provided for each. The validation set consists of four sub-folders (HV, PB, ME, and ML) deriving from a different source each, with a single-class (class of interest) annotation file provided for each audio file. The number of audio recordings in the validation set is 18 correspoding to about 6 hours. The training set data $T=( S_{I}, Y_{I} )|_{i=1}^\text{Ntrain}$ contains $S^{i}= \{S_{i} | Y_{i}= 1 \} \cup \{\widetilde{S_{l}}|Y_{i}= 0 \}$ and $Y_{i} = \{Y_{i}|Y_{i} \in \{0,1 \} \}$, where $\{S_{i} \}$ and $ \{\widetilde{S_{i}} \}$ are sets of positive and negative segments of class $i$, respectively. $\text{Ntrain}$ is the number of the total training segments. The evaluation set $E= ( S_{i}^{{}'}, Y_{i}^{{}'} )|_{i=1}^\text{Neval}$ contains $S_{i}^{{}'}= \{ S_{i}^{{}'}  \}$ and the label set $Y_{i}^{{}'}= \{Y_{i}^{{}'} \}$, where $\text{Neval}$ is the number of classes in the evaluation set. There is no overlap between the training set and validation set classes, which makes the dataset more challenging.

\subsection{Implementation Details}
The sampling rate of all audio is 22.5kHz. We first use the Short-Time Fourier transform with a window length of $1024$ and a hop size of $256$ to obtain a frequency-domain spectrogram. Then, by using Mel-scale filtering with 128 Mel bins, we have a 128-dimensional mel spectrogram. After that, we continue to extract the PCEN and MFCC features based on the mel spectrogram. During training, the model input was a stack of PCEN and MFCC features. If the length was less than $0.2$ seconds, it was padded with zeros. The initial learning rate was set to $0.001$, decayed by a factor of $0.65$ every $10$ epochs. Since there were only three types of bird sounds in the validation set, we used a 3-way 5-shot method for validation. If the validation accuracy did not improve for $10$ consecutive epochs, the model would stop training. The model with the best validation accuracy was used for evaluation. To better utilize the training data, we added a dynamic data loader to generate training data with random start times. In addition, we designed a post-processing strategy for sound categories based on the maximum length of positive events $t_{max} = \max [t_{1}, \cdots,t_{k} ]$. If the length of a detected positive event was less than $\alpha * t_{max}$ or greater than $\beta  * t_{max}$, we deleted it. During evaluation, we used $\beta = 2.0$, $\alpha = [ 0.1,0.2, \cdots, 0.9]$, and threshold values $h = [0.0,0.05, \cdots, 0.95]$. We calculated the PSD-ROC curve and PSDS using different combinations of $\beta $, $\alpha $, and $h$. We selected the best F-measure among all combinations of $\beta $,  $\alpha $, and $h$ as the final F-measure.

\subsection{Evaluation Criteria}

This paper uses two evaluation metrics. One is the F-measure, which is a balance between precision and recall. We use the F-measure as the main evaluation metric. The other metric used is the Polyphonic Sound Detection Score (PSDS)~\cite{Bilen}, which evaluates the algorithm's performance based on acoustic events. PSDS considers the occurrence, duration, and label of each acoustic event. Calculating the PSDS metric requires the use of forced alignment technology to align the events detected by the algorithm with the true events.

The specific steps for using the PSDS evaluation metric are as follows: First, preprocess the data, including data preprocessing and feature extraction. Second, segment the data to obtain information about the start and duration of each acoustic event. Then, evaluate the algorithm's detection performance, which includes event detection, event duration detection, and event label detection. For each event, if the algorithm detects the event, it is set to $1$, otherwise $0$. Evaluate whether the algorithm detects the duration of each acoustic event correctly. For each event, calculate the duration score based on the algorithm's detected duration and the true duration. Evaluate whether the algorithm correctly predicts the label. For each event, calculate the label score based on the algorithm's predicted label and the true label. For each acoustic event, calculate the weighted average of its occurrence score, duration score, and label score to obtain its PSDS score. Average the PSDS scores of all acoustic events to obtain the algorithm's average PSDS score. Using the PSDS evaluation metric can objectively evaluate the performance of bioacoustic event detection algorithms and adjust the evaluation for different types of acoustic events to improve the accuracy and reliability of the evaluation. In PSDS, the detection tolerance (DTC) is set to $0.5$, the ground truth (GTC) is set to $0.5$, and the maximum effective false alarm rate is $100$. Since we do not perform multi-channel detection, we do not use the cross-triggering tolerance criteria (CTTC).
\begin{figure}
    \centering
    \includegraphics[width=0.9\linewidth]{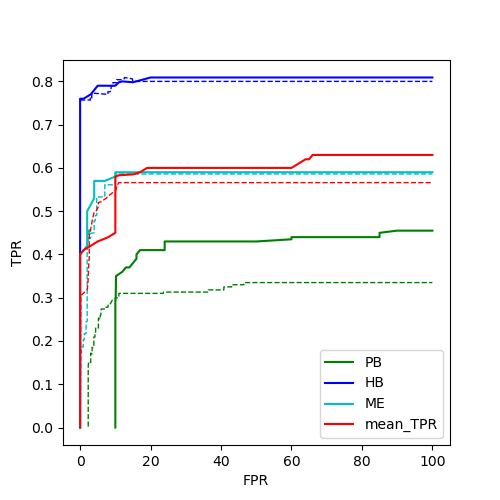}
    \caption{Polyphonic Sound Detection Score (PSDS) on the dataset of task 5 of DCASE 2022, where the green color, blue color, and cyan color represent the results on the PB type, HB type, and ME type of bird sounds, respectively. The red color corresponds to $\text{mean}_\text{TPR}$. The solid line corresponds to the results of the proposed MCS-Net, and the dashed line corresponds to the result of the Byte Dance team. TPR and TPR are the true positive rate and the false positive rate, respectively. The PSDS values are between zero and one, where better-performing systems achieve higher scores.}
    \label{fig:2}
\end{figure}
\begin{figure}
    \centering
    \includegraphics[width=\linewidth]{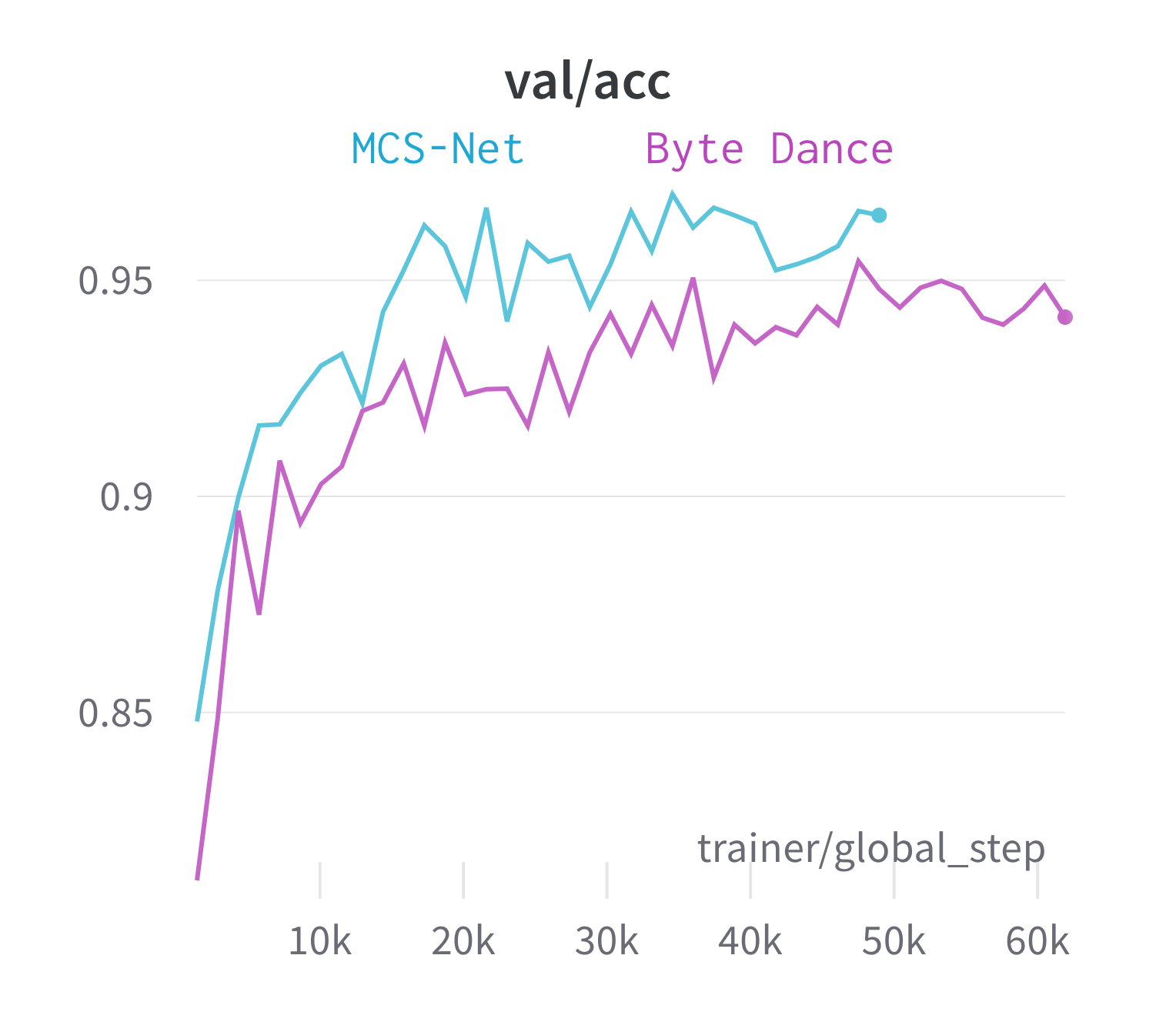}
    \caption{MCS-Net model and Byte Dance model validation comparison chart, where the horizontal axis is the training step and the vertical axis is the accuracy, the cyan line in the chart represents the MCS-Net model and the purple line represents the Byte Dance model, the higher the accuracy the better, from the chart you can clearly see that the cyan line is higher than the purple line.}
    \label{fig:3}
\end{figure}
\subsection{Results}
The performance of the proposed MCS-Net model on the evaluation set is reported in Table~\ref{table:1}. In the table, we compare our results with previous works, including the results of the Template Matching method from the DCASE 2021 competition, the ProtoNet baseline model, and the Byte Dance team, who won the DCASE 2022 Take5 competition with an F-measure score of $62.73\%$ and PSDS of $57.52\%$. Our proposed MCS-Net model achieved an F-measure score of $66.84\%$ and PSDS of $58.98\%$, outperforming the previous methods.

As shown in Fig.~\ref{fig:2}, the ROC curve shows that the PB class is the most difficult to detect, the HB class is the easiest to detect, and the ME class achieves average performance on the evaluation set. By comparing the two sides of the curve, it can be seen that the MCS-Net model has significantly improved the detection of the most difficult PB class.

As shown in Fig.~\ref{fig:3}, the purple line represents the Byte Dance model, and the blue line represents our proposed MCS-Net model. The improved MCS-Net model completed training in 1 hour 19 minutes and 27 seconds with a precision of 97\%, which is faster and more accurate than the Byte Dance model with fewer steps.

\section{Conclusions}
In this paper, we introduce the MCS-Net network model, which employs prototype network and metric learning techniques in the context of limited data samples. By incorporating residual blocks and Channel Spatial SE blocks into the prototype network, the MCS-Net model enhances network stability and focuses on channels and spatial dimensions, thus improving feature representation capabilities. The MCS-Net model demonstrates strong learning and generalization abilities even with small sample data. On the publicly available DCASE2022-Take5 dataset, the MCS-Net model achieved an F-measure score of $66.84\%$ and a PSDS of $58.98\%$, representing absolute improvements of $4.11\%$ and $1.46\%$ in F-measure and PSDS, respectively, compared to previous works. These results showcase the effectiveness of the MCS-Net model in enhancing bird sound classification and detection performance.

\nocite{*}
\bibliographystyle{IEEEtran}
\bibliography{reference}
\end{document}